\documentclass[10p,times]{article}

\usepackage{geometry}
\usepackage{amssymb}
\usepackage[figuresright]{rotating}
\usepackage{chngcntr}
\usepackage{amsmath}
\usepackage{epsfig}
\usepackage{epstopdf}
\usepackage[T1]{fontenc}
\usepackage{amssymb}
\usepackage{amsmath}
\usepackage{amsfonts}
\usepackage{verbatim}
\usepackage{graphicx}
\usepackage{hyperref}
\usepackage{array}

\usepackage{placeins}

\usepackage{tikz}
\usetikzlibrary{positioning}
  \usepackage[ruled]{algorithm}
\usepackage{algpseudocode}
\usepackage{lipsum}
\usepackage{multicol}
\usepackage{float}
\usepackage{caption}
\usepackage{subcaption}

\begin{document}
\title{A Novel Reconfigurable Hardware Design for Speech Enhancement Based on Multi-Band Spectral Subtraction Involving Magnitude and Phase Components}

\author
{Tanmay Biswas$^{1}$, Sudhindu Bikash Mandal$^{1}$, Debasree Saha$^{2}$, Amlan Chakrabarti$^{3}$\\
\\
\normalsize{IEEE Student Member$^{1}$, IEEE Member$^{2}$, IEEE Senior Member$^{3}$}\\
\normalsize{A.K.Choudhuri School Of Information Technology, University Of Calcutta$^{1,2,3}$}\\
\normalsize{(tanmay123g, sudhindu.mandal)@gmail.com$^{1}$, debasri\_cu@yahoo.co.in$^{2}$, acakcs@caluniv.ac.in$^{3}$}\\
}

\maketitle

\begin{abstract}
This paper proposes an efficient reconfigurable hardware design for speech enhancement based on multi band spectral subtraction algorithm and involving both magnitude and phase components. Our proposed design is novel as it estimates environmental noise from speech adaptively utilizing both magnitude and phase components of the speech spectrum. We performed multi-band spectral subtraction by dividing the noisy speech spectrum into different non-uniform frequency bands having varying signal to noise ratio (SNR) and subtracting the estimated noise from each of these frequency bands. This results to the elimination of noise from both high SNR and low SNR signal components for all the frequency bands. We have coined our proposed speech enhancement technique as Multi Band Magnitude Phase Spectral Subtraction (MBMPSS). The magnitude and phase operations are executed concurrently exploiting the parallel logic blocks of Field Programmable Gate Array (FPGA), thus increasing the throughput of the system to a great extent. We have implemented our design on Spartan6 Lx45 FPGA and presented the implementation result in terms of resource utilization and delay information for the different blocks of our design. To the best of our best knowledge, this is a new type of  design for speech enhancement application and also a first of its kind implementation on reconfigurable hardware. We have used benchmark audio data for the evaluation of the proposed hardware and the experimental results show that our hardware shows a better SNR value compared to the existing state of the art research works.\\

\end{abstract}

Keyword's: Spectral Subtraction, Multi Band Spectral Subtraction, Digital Signal Processing (DSP), Field Programmable Gate Array (FPGA), System Generator.

\section{Introduction}

Speech enhancement aims to improve the quality of speech in a noisy environment. The spectral subtraction technique is a well known technique for speech noise elimination, which was originally introduced by S. Boll~\cite{Boll:spectral_subtraction}. An upgraded version was introduced by Berouti et al.~\cite{Berouti:better_version} for the musical noise reduction. The general principle behind the spectral subtraction is to estimate noise from the magnitude spectrum, which then gets subtracted from the original signal keeping the phase part of the spectrum unchanged. This general spectral subtraction technique results to three kinds of error~\cite{Zhang:real_imaginary} viz. error in noise estimation, error due to ignoring the speech-noise cross term in magnitude spectrum and error due to noisy phase spectrum with clear magnitude spectrum in signal reconstruction. The performance of speech enhancement is put down due to these errors. These errors has been reported in~\cite{Lin_2003}~\cite{evans}~\cite{Lu_2008} for speech enhancement and speech recognition methods. When SNR of the signal is high, the noisy phase is close to the clean phase and the above methods work properly. But, when SNR drops then the cross term errors are produced, and the phase of the noisy signal plays the more seeming role in the clean magnitude signals and affects the reconstruction process. Recently, real and imaginary modulation spectral subtraction for speech enhancement was introduced by Yi Zhang et al.~\cite{Zhang:real_imaginary}, where the subtraction procedure performed on both real and imaginary parts of the spectrum. Also the real world noise affects signal in various time intervals, which is also called colored noise. The real world noise spectrum are not flat like white Gaussian noise. The multi band spectral subtraction method for speech enhancement was introduced by Kamath~\cite{kamnath}, where the spectrum was divided into several bands for efficient noise reduction. In ~\cite{multi2}, design of multi band spectral subtraction was proposed based on the magnitude compensation and phase modification. In~\cite{phase_based}, we can find a phase based dual microphone algorithm for robust speech enhancement. Plenty of research work based on spectral subtraction algorithm can be found in~\cite{IET}~\cite{ref:2013a}~\cite{wiener_2006}~\cite{Ephriam:mmse}~\cite{mallat:blockthresholding}
~\cite{wavelet:2006}~\cite{els_10}. Speech enhancement based on hardware software co-design using FPGA  platform can be found in~\cite{ref:2013}~\cite{design1}~\cite{design2} ~\cite{design3}. 

\par
In this paper, we have performed noise estimation from both magnitude and phase spectrum by dividing the whole noisy speech spectrum into different non-uniform linearly shaped frequency bands and then subtracted the estimated noise from each frequency bands with different SNR over subtraction factor values  ${\alpha}$. This correctly justifies that our proposed hardware design for speech enhancement performs well for both high SNR signal as well as low SNR signal with different frequency bands. The hardware execution can be carried out in two ways: (a) off the shelf Digital Signal Processors (DSPs) and (b) FPGAs. We have chosen FPGA as our target hardware as it gives the opportunity of parallel computing involving the configurable logic cells~\cite{DSP:2010} and dedicated DSP blocks. This leads to faster execution of hardware tasks, satisfying our primary objective. We have used the Xilinx System Generator tool in the MATLAB/SIMULINK environment~\cite{matlab:verification} to design and verify our hardware. Here, we convey the comparative experimental results of SNR performance of the proposed architecture against the Magnitude Spectral Subtraction (MSS)~\cite{Boll:spectral_subtraction}, Magnitude Phase Spectral Subtraction (MPSS)~\cite{Zhang:real_imaginary} and Multi Band Magnitude Spectral Subtraction (MBMSS)~\cite{kamnath} for different noisy signals, which clearly infers that our design yields better performance. We also convey the resource utilization and delay information of the proposed architecture. The major contributions of this work can be summarized as follows:

\begin{enumerate}
\item Proposal of a new speech enhancement method, relatively more robust as compared to the state of the art works (MSS~\cite{Boll:spectral_subtraction}, MPSS~\cite{Zhang:real_imaginary}, MBMSS~\cite{kamnath}), this is indicated by the improved performance in terms of SNR.
\item FPGA based hardware design and implementation of the proposed speech enhancement methodology.

\end{enumerate}

 This paper is organized as follows. In section $2$, a brief background of magnitude spectral subtraction and multi band spectral subtraction algorithms are presented; our proposed hardware design for speech enhancement architecture is presented in Section $3$; in Section $4$ hardware implementation and in Section $5$ performance analysis are presented; concluedary remarks in Section $6$.

\section{Background}
In this section we discuss some of the fundamental issues related to spectral subtraction technique that are extremely important to understand our work presented in the next subsequent sections.

\subsection{Spectral Subtraction Algorithm}

Spectral subtraction is a procedure for restoration of the power spectrum or the magnitude spectrum of a signal observed in additive noise, through subtraction of an estimate of the average noise spectrum from the noisy signal spectrum. The noisy signal in time domain is represented as:
\begin{equation}
y(m)= x(m) + n(m)                                                
\end{equation}
where $y(m)$, $x(m)$ and $n(m)$ are the signal, additive noise and the noisy signal respectively and $m$ is the discrete time index.
\\The frequency domain noisy signal model corresponding to equation (1) can be represented as:
\begin{equation}
 Y(f) = X(f) + N(f)                                                      
\end{equation}
 Where $Y(f)$, $X(f)$ and $N(f)$ are the frequency domain signals corresponding to $y(m)$, $x(m)$ and $n(m)$ respectively.
 \\The noise estimation filter calculates $N(f)$ from the noisy spectrum. The magnitude of $N(f)$ is calculated by its average value during non speech activity. Spectral error~\cite{Ephriam:mmse} comes from subtraction estimator. It reduces by simple modification like magnitude averaging, half wave rectification, residual noise reduction and additional signal attenuation during non speech activity.
 \\The discontinuities  at the end point of the segment can be done by windowing of the signal and can be expressed as:
 \begin{equation}
y_{\omega}(m) = x_{\omega}(m) + n_{\omega}(m). 
  \end{equation}
  Windowing signal can be expressed in frequency domain as:
  \begin{equation}
Y_{\omega}(f) = W(f) * Y(f)
                       =  X_{\omega}(f) + N_{\omega}(f)
  \end{equation}
  where the operator * denotes convolution.
  \\ A scaled estimate of the magnitude spectra of the noise signal $\hat{N}_{\omega}(f)$ is subtracted from the corresponding spectra of the noisy signal $Y_{\omega}(f)$ to estimate the clean voice $\hat{S}_{\omega}(f)$ ,
  \begin{equation}
|\hat{S}_{\omega}(f)|^{\gamma} = |Y_{\omega}(f)|^{\gamma} - |{\alpha}\hat{N}_{\omega}(f)|^{\gamma}
\end{equation}
Noise signal is estimated and the frequency dependent subtraction factor ${\alpha}$ is included to compensate the overestimation of the instantaneous noise spectrum. ${\gamma}=1$ for the magnitude spectral subtraction and ${\gamma}=2$ for power spectral subtraction.
The enhanced signal spectrum is obtained using the magnitude estimate $\hat{S}(f)$ and phase $\phi(f)$ of the corrupted input signal,

\begin{equation}
\hat{S}(f) = |\hat{s}f|e^{j\phi(f)}
\end{equation}

Finally, the clean signal is obtained by the inverse fourier transform of $\hat{S}(f)$,

\begin{equation}
\hat{s}(m) =  F^{-1}{\hat{S}(f)}
\end{equation}

This general spectral subtraction method provides better results of the speech enhancement for high SNR signals compared to the low SNR signals. A combination of magnitude and phase spectral subtraction methods provides speech enhancement of both high and low SNR signals~\cite{Zhang:real_imaginary}.

\subsection{ Multi Band Spectral Subtraction Algorithm}

\begin{figure}[!htb]
\centering
\hspace*{1pt}
\includegraphics[height= 6 cm, width=13 cm]{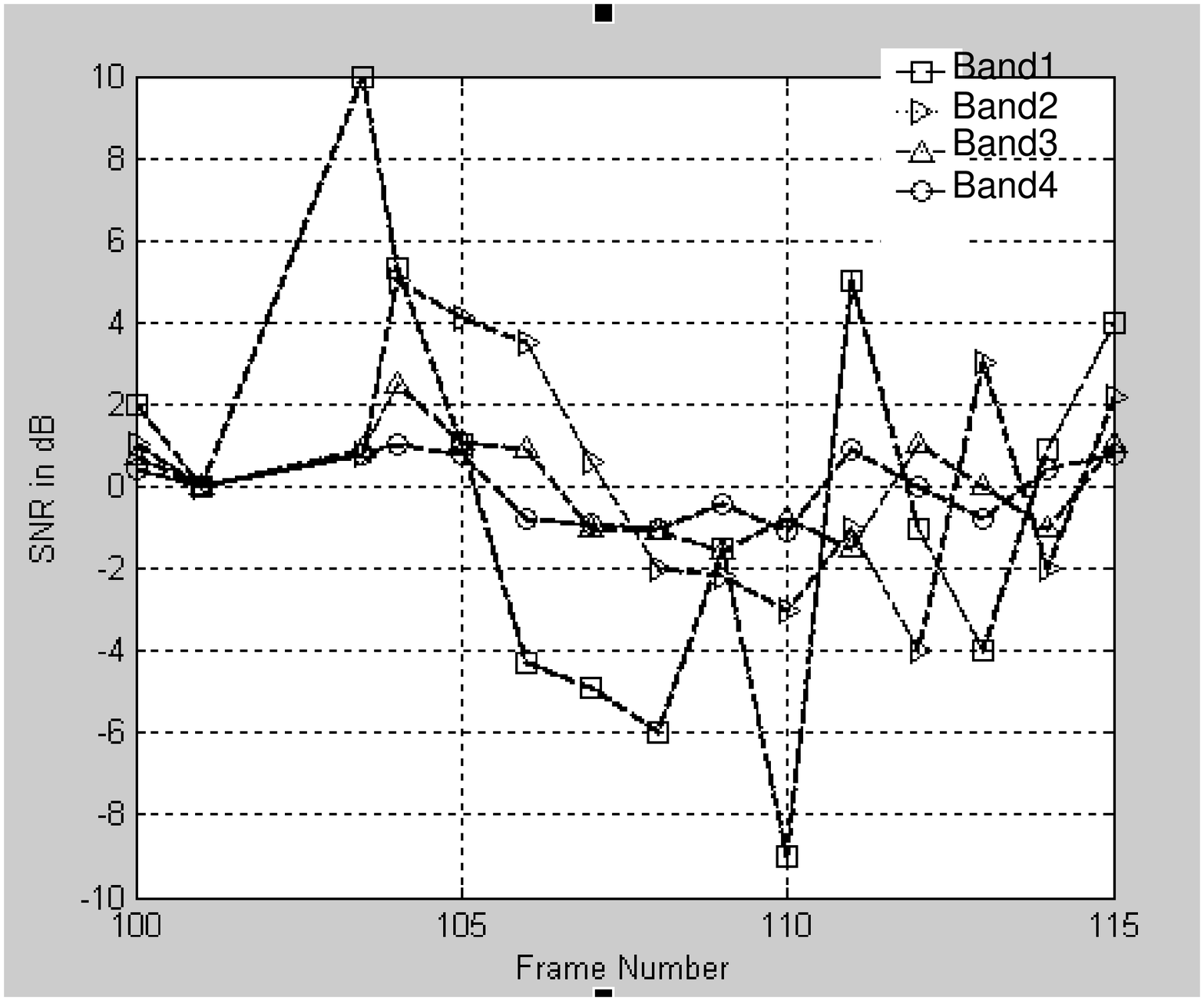}
\vspace*{-8pt}
\caption{Segmental SNR comparison for different frequency bands}
\label{fig1}
\end{figure}

Most of the real world noise are colored noise, which affects the signal at various time interval. Multi band spectral subtraction algorithm~\cite{kamnath} provides subtraction over individual frequency bands for the better speech enhancement.  
\\ The clean speech spectrum of equation (2) can be represented as:
  \begin{equation}
|\hat{S}_{\omega}(f)|^{\gamma} = |Y_{\omega}(f)|^{\gamma} - {\alpha}|\hat{N}_{\omega}(f)|^{\gamma}
\end{equation}
where $\alpha$ is the over subtraction factor, which is the function of the segmental SNR. General spectral subtraction methods assume that the noise is affected uniformly and  the over subtraction factor $\alpha$ is subtracted over the whole spectrum. In real world the noise is effected in random phenomenon. The colored noise affects the signal spectrum differently at various frequencies. So the segmental SNR values change at different frequency bands~\cite{kamnath}. The change of estimated SNR value for four frequency bands is shown in Fig.~\ref{fig1}. This four frequency bands are linearly spaced.

The speech spectrum is divided into different non overlapping bands and the subtraction procedure is done over each band independently. So, the enhanced signal spectrum of the $i$th frequency bands is,  
  \begin{equation}
|\hat{S}_i{\omega}(f)|^{\gamma} = |Y_i{\omega}(f)|^{\gamma} - {\alpha_i}{\delta_i}|\hat{N}_i{\omega}(f)|^{\gamma}  
\end{equation}
 where $\alpha_i$ is the over subtraction factor of the $i$th frequency band and $\delta_i$ is the tweaking factor of each $i$th band. $b_i$ and $e_i$ are the beginning and ending frequency bins of $i$th frequency band. The over subtraction factor $\alpha$ is directly depended on the segmental SNR of the signal and is calculated as:
   \begin{equation}
SNR_i(db) = 10 *log_{10} \sum\limits_{f=b_i}^{e_i}(|Y_i{\omega}(f)|/|N_i{\omega}(f)|)^2
\end{equation}

Depending upon the $SNR_i$ value the over subtraction factor $\alpha_i$ evaluated as:
 \begin{equation}
\alpha_i = \left\{ \begin{array}{rl}
 5 &\mbox{ $SNR_i<5$} \\
 4-3/20(SNR_i) &\mbox{$-5<SNR_i<5$}\\
 1 &\mbox{ $SNR_i>20$}
       \end{array} \right. 
\end{equation}

The over subtraction factor has a control in subtraction for each of the frequency bands. The subtraction factor $\delta_i$ provides an additional degree of control for each frequency bands.The value of $\delta_i$~\cite{kamnath} is specified by the following equation:
  \begin{equation}
\delta_i = \left\{ \begin{array}{rl}
 1 &\mbox{ $f_i<1KH_z$} \\
 2.5 &\mbox{$1KH_z<f_i<FS/2-2 KH_z$}\\
 1.5 &\mbox{ $f_i>FS/2-2 KH_z$}
       \end{array} \right. 
\end{equation}
Where $f_i$ is the upper frequency band and $FS$ is the sampling frequency.

After the subtraction of the estimated noise from all the frequency bands with different segmental SNR, we create the enhanced frequency bands. 

\section{ Proposed Hardware Design}
From the above discussion we observe that the MSS technique enhances the high SNR signals, MBMSS technique enhances the high SNR signals for different frequency bands. We propose a novel MPMBSS technique, which enhances both high SNR and low SNR signals for the different frequency bands.

\begin{figure*}[!htb]
\centering
\hspace*{0cm}
\includegraphics[height=12 cm, width=17cm] {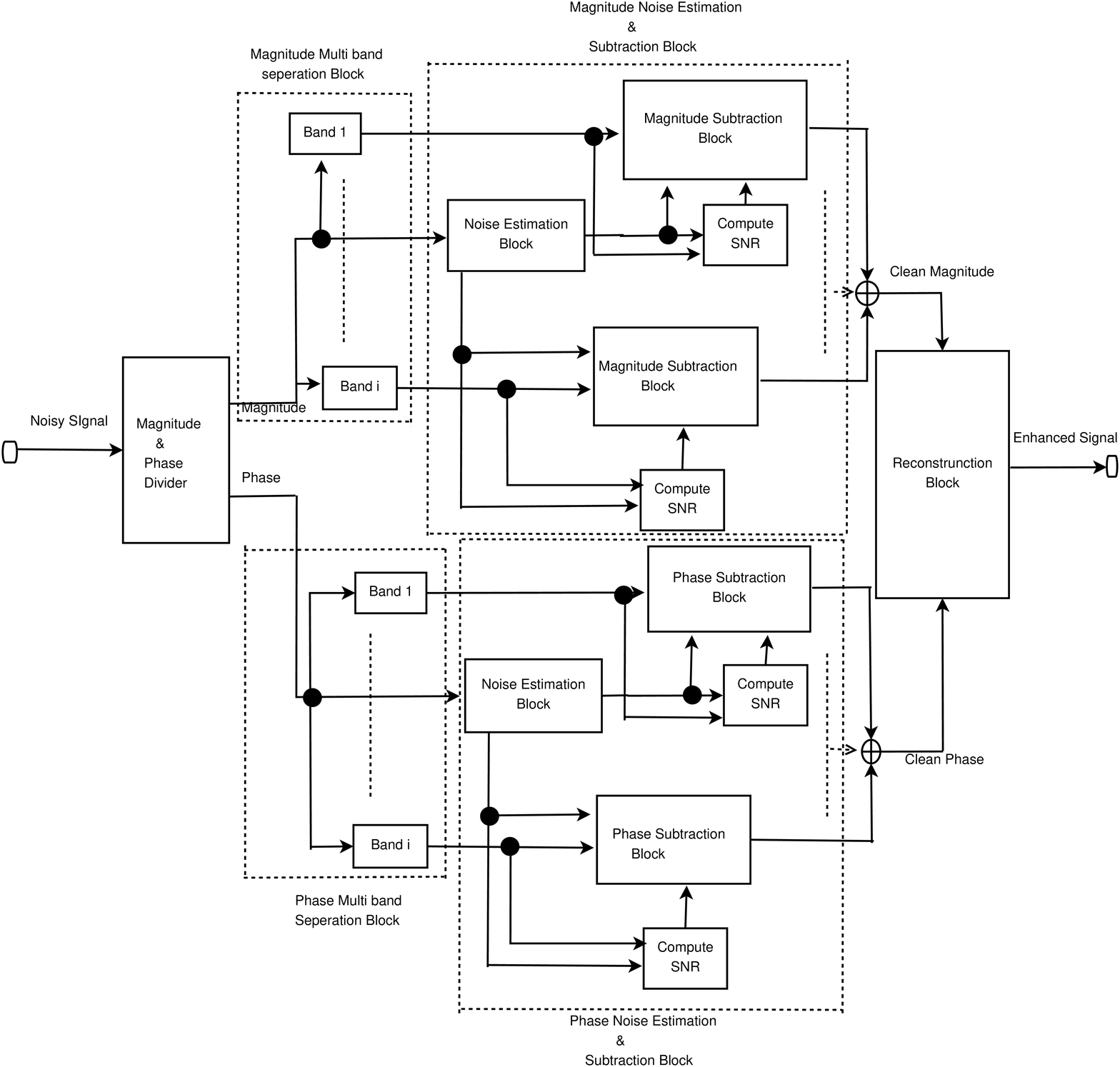}
\caption{Proposed Architecture}
\label{fig2}
\end{figure*}

In our proposed design we have four principle blocks namely magnitude multi band separation block, magnitude noise estimation-subtraction block, phase multi band separation block and phase noise estimation-subtraction block. The magnitude and phase operations are executed in parallel. The proposed architecture is shown in Fig.~\ref{fig2}.

 From equation (9) we have,
 \begin{equation}
|\hat{S}_i{\omega}(f)|^{\gamma} = |Y_i{\omega}(f)|^{\gamma} - {\alpha_i}{\delta_i}|\hat{N}_i{\omega}(f)|^{\gamma}  
\end{equation}

Where $\alpha_i$ is the over subtraction factor of the $i^{th}$ frequency bands. The magnitude and the phase spectrum of the signal is divided into different frequency bands and the subtraction is done on each frequency bands from the estimated noise of both magnitude and phase spectrum of the signal.
\\ The noise estimated from the magnitude and noise spectrum of the signal are $\hat{N}_i{\omega}mg(f)$ and$\hat{N}_i{\omega}ph(f)$ respectively and it is subtracted from the noisy magnitude and phase spectrum with different frequency bands depending upon the $\alpha_i$.
 \begin{equation}
|\hat{S}_i{\omega}mg(f)|^{\gamma} = |Y_i{\omega}mg(f)|^{\gamma} - {\alpha_i}{\delta_i}|\hat{N}_i{\omega}mg(f)|^{\gamma}  
\end{equation}

 \begin{equation}
|\hat{S}_i{\omega}ph(f)|^{\gamma} = |Y_i{\omega}ph(f)|^{\gamma} - {\alpha_i}{\delta_i}|\hat{N}_i{\omega}ph(f)|^{\gamma}  
\end{equation}

Where $|\hat{S}_i{\omega}mg(f)|$ and $\hat{S}_i{\omega}ph(f)$ are the clean magnitude and phase spectrum of the $i^{th}$ frequency bands respectively. $|Y_i{\omega}mg(f)|$ and $Y_i{\omega}ph(f)$ are the noisy magnitude and phase spectrum of the ${i^{th}}$ frequency bands respectively. And $\alpha_i$ is calculated from the compute SNR block.

The enhanced magnitude spectrum $|\hat{S}_i{\omega}mg(f)|$ and the enhanced phase spectrum $|\hat{S}_i{\omega}ph(f)|$ of the signal are combined together at the time of reconstruction of the signal.  

\begin{equation}
\hat{S}_mp(f) =  F^{-1}{\hat{S}_mp(f)}
\end{equation} 

$\hat{S}_mp(f)$ is the enhanced speech signal of the noisy signal.
\\We claim that the proposed design has an increased signal to noise ratio (SNR) as compared to the other existing state of the art architectures. The magnitude and phase operations are executed concurrently exploiting the parallel logic blocks of field programmable gate array (FPGA). 

\section{ Hardware Implementation}

\begin{figure*}[!htb]
\centering
\hspace*{0cm}
\includegraphics[height=11 cm, width=17cm] {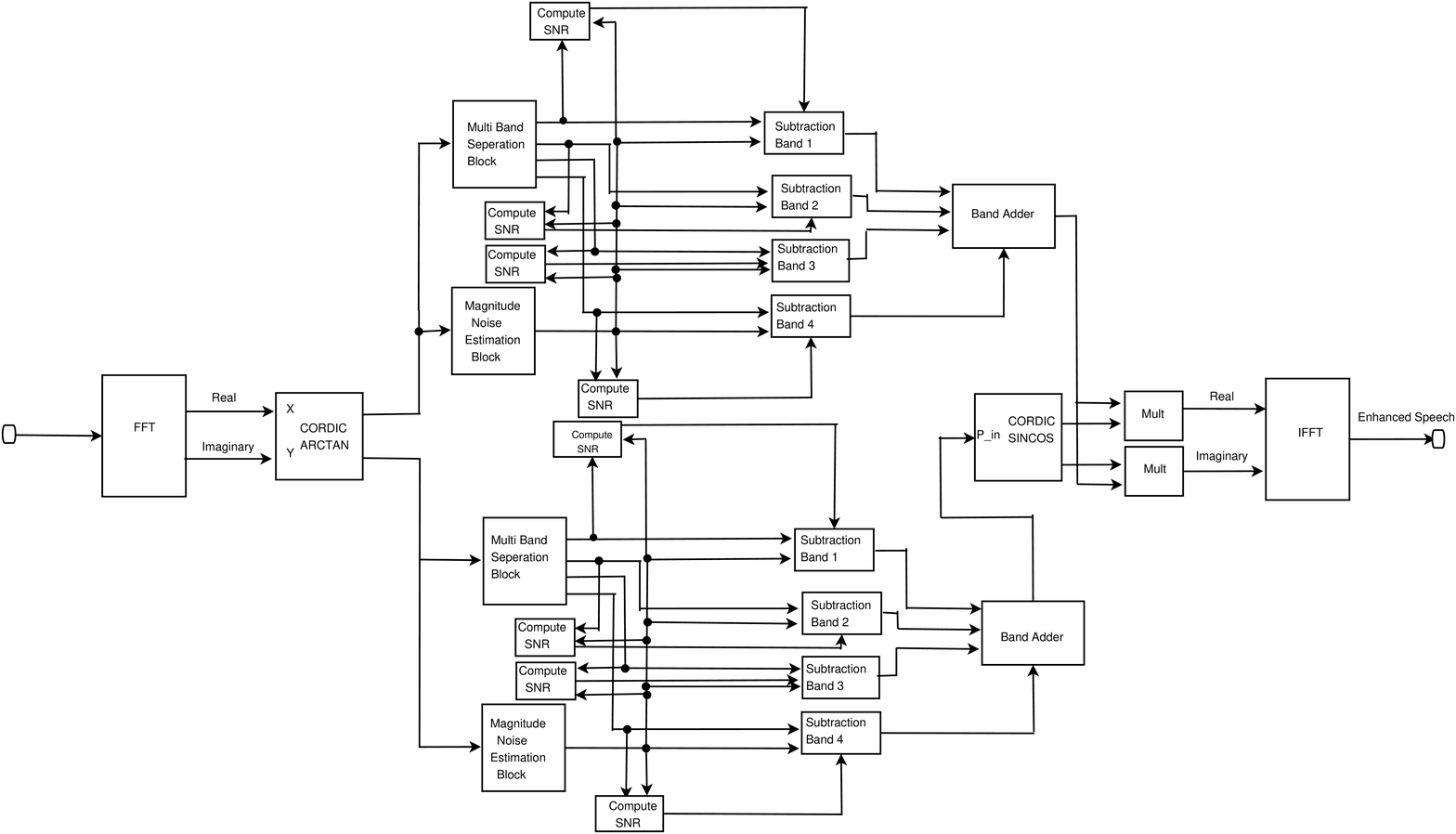}
\caption{Hardware Design}
\label{fig3}
\end{figure*}

The proposed architecture is implemented on the reconfigurable FPGA hardware. The time domain noisy speech signal is converted into the frequency domain signal by the Fast Fourier Transform (FFT) block. This signal is divided into magnitude and phase components using CORDIC ARCTAN DSP block. From the magnitude and phase spectrum, noise is estimated by the noise estimation block. The magnitude and  phase spectrum are divided into four frequency bands~\cite{kamnath} by multi band separation block and over subtraction factor ($\alpha_i$) is calculated from each of the bands by using the SNR computation block. Magnitude and phase subtraction block subtracts the estimated noise from each of the frequency bands with different $\alpha_i$. Enhanced magnitude and phase spectrum of the different frequency bands are combined together by the adder block and thus generating the enhanced magnitude and phase spectrum of the signal. 
\\ The enhanced phase signal passes through the CORDIC SINCOS to generate the real and imaginary form of the phase spectrum. These real and imaginary phases are combined with the enhanced magnitude spectrum using the multiplier block and is passed through the inverse FFT (IFFT) block to reconstruct the signal. Output of the IFFT block gets the enhanced speech signal. The block diagram of the proposed hardware design is shown in Fig.~\ref{fig3}.

\subsection{Fast Fourier Transform}
\begin{figure}
    \centering
    \begin{subfigure}[b]{0.3\textwidth}
        \centering
        \includegraphics[height= 5 cm, width=1.5\textwidth]{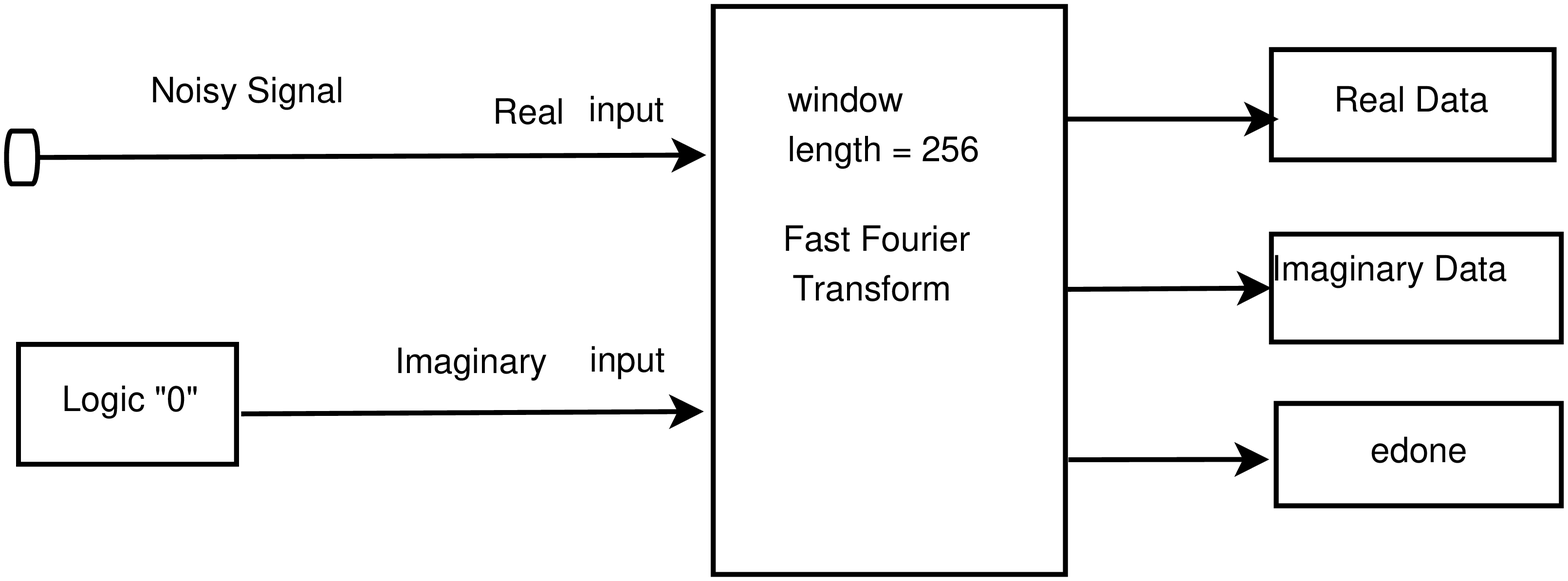}
        \caption{FFT Block}
      \label{fig4a}
    \end{subfigure}
    \hfill
    \begin{subfigure}[b]{0.4\textwidth}
        \centering
        \includegraphics[height= 5 cm, width=1\textwidth]{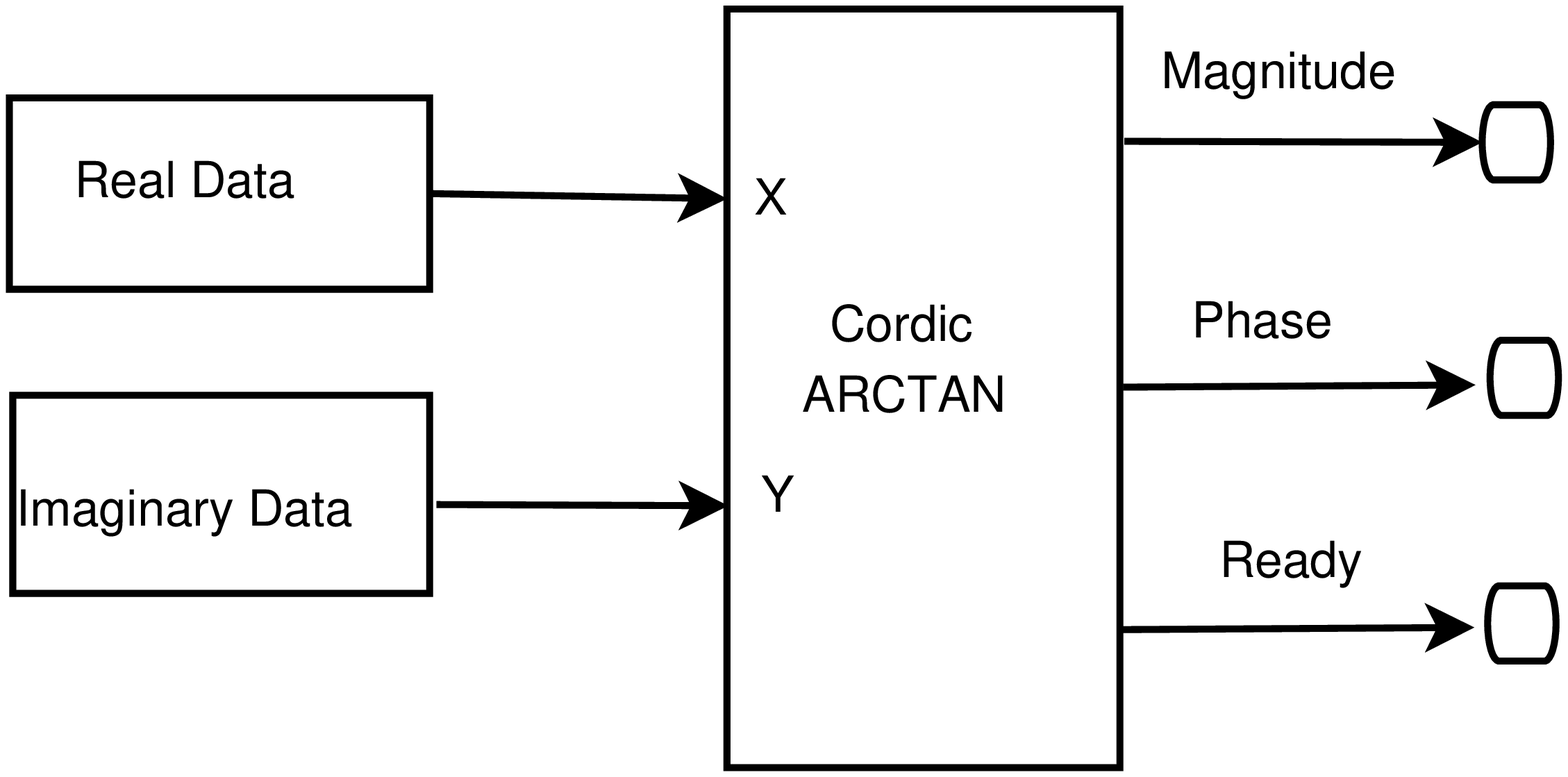}
        \caption{Magnitude and Phase Separation}
       \label{fig4b}
    \end{subfigure}
 \caption{Block Diagram of FFT and CORDIC ARCTAN}
    \label{fig4}
\end{figure}

The real world noisy speech signals are in the time domain signals. To convert this time domain signal to frequency domain we require the fourier transform over the signal. Fast Fourier Transform (FFT) block is used to perform the fourier operation on the signal. The FFT block generates the real as well as the imaginary components of the signal. We have used the FFT block of the Xilinx system generator platform. The option input/output was chosen for the FFT to implement its pipelined versions. For the performance optimization of the FFT block 4 multiplier structures are used and the phase factor is set to 8. Also the signal is segmented on non overlapping window of 256 samples. The data is recorded in the 2 stages of the block RAM (BRAM). The interface of the FFT block is shown in Fig.~\ref{fig4a}. The real and imaginary components are transfered to the magnitude and phase separation block.

\subsection{Magnitude and Phase Separation Block}
Output of the FFT block drives the CORDIC ARCTAN block where the real and imaginary signal are divided into their magnitude and phase format. The magnitude $Y_{\omega}mg(f)$ and the phase $Y_{\omega}ph(f)$ are passed through the magnitude noise estimation block and phase noise estimation block respectively. 

In the cordic ARCTAN block~\cite{sysgen}, architectural configuration is set to parallel mode for high throughput. The pipeline mode is set to maximum and the phase format is set to radians mode. The output width of this block is set to 16. The output pins are magnitude, phase and ready pin of the Cordic ARCTAN block. The block diagram of the magnitude and phase separation block is shown in Fig.~\ref{fig4b}.

The magnitude and phase spectrum are fed to the magnitude noise estimation block and phase noise estimation block respectively in parallel, where the band separation, noise estimation and subtraction processes are done.

\subsection{Noise Estimation Block}
\begin{figure}
    \centering
    \begin{subfigure}[b]{0.3\textwidth}
        \centering
        \includegraphics[height= 7 cm, width=1.5\textwidth]{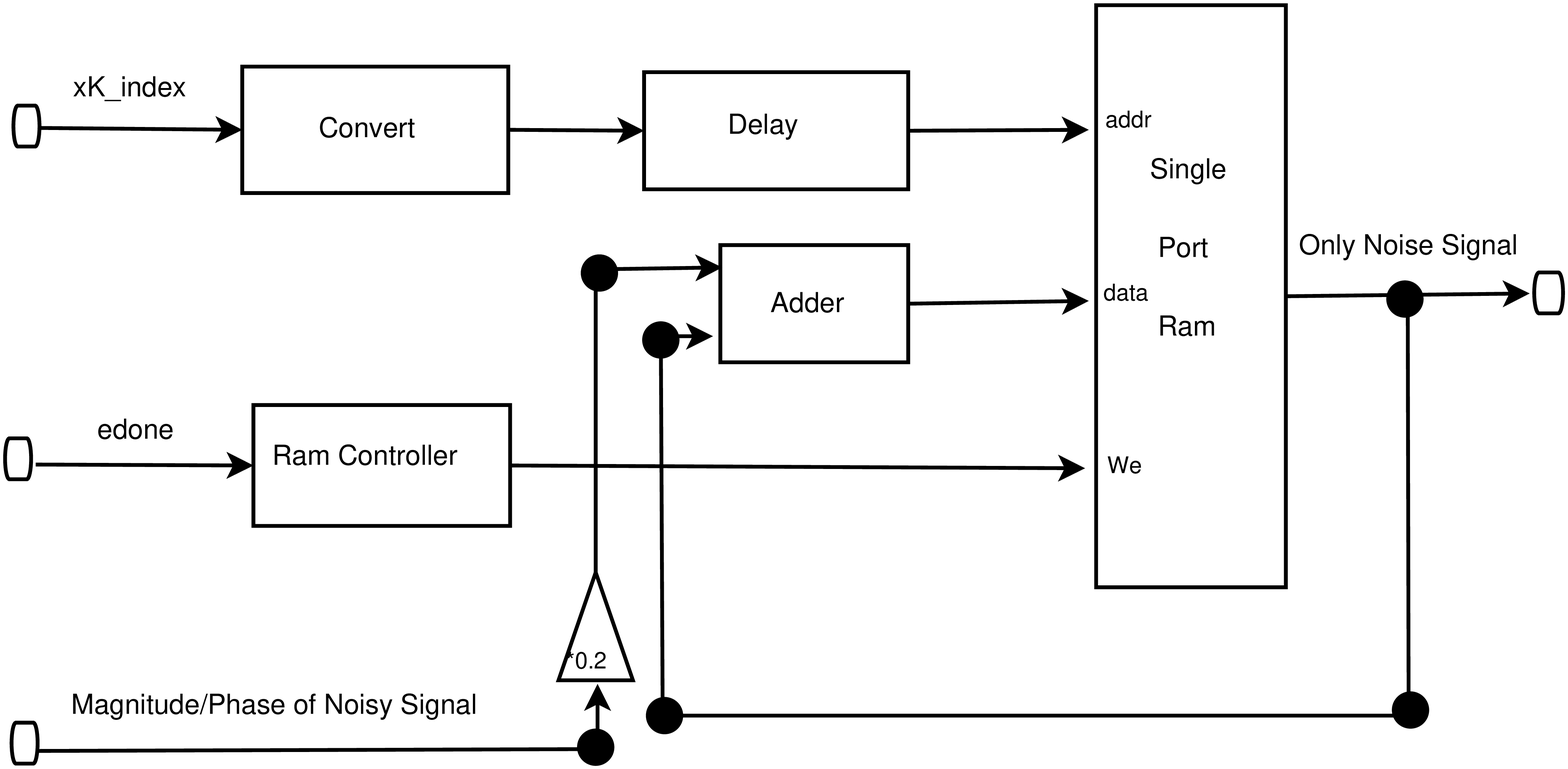}
        \caption{Magnitude/Phase Noise Estimation Block}
      \label{fig5a}
    \end{subfigure}
    \hfill
    \begin{subfigure}[b]{0.5\textwidth}
        \centering
        \includegraphics[height= 7 cm, width=1\textwidth]{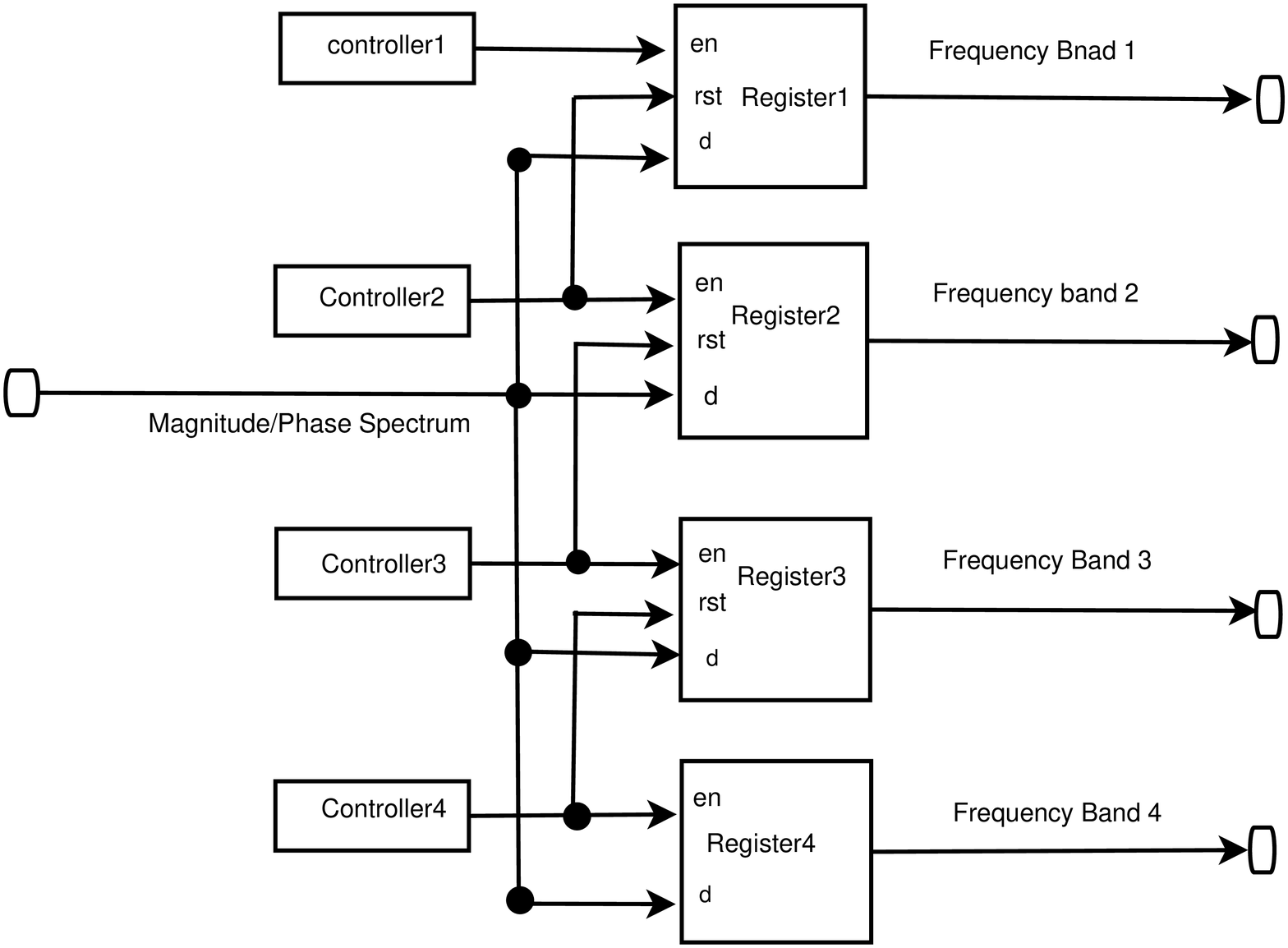}
        \caption{Multi Band Separation Block}
       \label{fig5b}
    \end{subfigure}
 \caption{Block Diagram of Noise Estimation and Multi Band Separation Block}
    \label{fig5}
\end{figure}

In the noise estimation blocks, noise is estimated during the first few samples where noise is only present. Our design is adaptive in nature with the only constraint that a few initial samples of the input signal for a duration of $1.25ms$ is only noise, which is a fair constraint for speech communication. The block diagram of the magnitude and phase noise estimation block is shown in Fig.~\ref{fig5a}.

Magnitude spectrum and phase spectrum of the signal are passed to the magnitude noise estimation block and phase noise estimation block respectively. In this design, one single port RAM is used to calculate noise power. RAM is set in write before read mode. Write enable pin of the single port RAM is active only for initial few samples (here 5 samples), which is taken care by the RAM controller block. One counter and one relational block are used to architect the RAM controller block. The first 5 samples of data are written on the RAM block and the other next samples are passed through the RAM block in the read mode. The noise samples that are stored in the RAM block are ready to be subtracted from the different frequency bands having magnitude spectrum and phase spectrum.


\subsection{Multi Band Separation Block} 
The magnitude spectrum and phase spectrum of the signal also pass through multi band separation block. The magnitude spectrum and phase spectrum is divided into four frequency bands, which are linearly spaced. Four registers store the sample value of the four frequency bands until the reset signals are disabled. The multi band separation block is shown in Fig.~\ref{fig5b}.

There are four controller blocks (Fig.~\ref{fig5b}) that are used to divide the signal linearly into four frequency bands of the single signal. Each controller is connected to the enable pin of the registers. The reset pins of the first three register are enabled when the immediate next register enable pins are enabled. The fourth register have no reset pin, so the fourth register carry the rest of the signal. When the signal are in the first register by the controller1, the reset pin is disabled and the sample signal values are passed through the register and falls into the first subtraction block where the subtraction procedure is done. When controller2 becomes active then register2 is ready to accept signal and also register1 is in the reset mode by  controller2 to avoid override the signals. When the register4 becomes active then rest of the signal passes through this register. If we require more frequency bands, then we need to connect more registers and controllers. But, we investigate in our design that four frequency bands of the signal give better throughout. This architecture is modeled in parallel configuration.
\\After subtraction of each bands from the estimated noise spectrum we reconstruct the signal into a single band frequency. Here we used multi band adder block to reconstruct the four signals. But before addition of the four enhanced signals, same controllers and registers which have been used for the separation are used for the same time format of the original signals.

\subsection{Signal to Noise Ratio Computation} 
For the proposed architecture over subtraction factor $\alpha$ may vary for different frequency bands and depends on the signal to noise ratio for each frequency bands. The variation of $\alpha$ is described in equation (11). The architecture of calculating SNR is shown in Fig.~\ref{fig6a}.

Maximum amplitude of signal and noise are calculated by the relational block, multiplexer and register block. The enable pin of the register is connected to the output of the controller to keep the same bandwidth of the divided signal. The controller blocks are the same which was used previously to separate frequency bands. Those four controllers are used to the SNR computation block to get the maximum amplitude of the signal and noise. The output of the multiplexer holds the maximum value for each sample comparison. The registers pass the final maximum value of the signal samples by the controller block. Maximum amplitude of signal and noise pass through a division block to get the SNR value. In our proposed architecture, we require four SNR computation block for magnitude spectrum and four for the phase spectrum to estimate SNR values. The over subtraction factor $\alpha_i$ of each bands are calculated using the corresponding SNR values. Then subtraction done between each frequency bands and estimated noise with the different $\alpha_i$.

\subsection{Subtraction Block}

In the subtraction block, estimated noise spectrum with over subtraction factor is subtracted from the signal spectrum of each of the frequency bands. The enhanced spectrum of $ith$ frequency band is,

 \begin{equation}
|\hat{S}_i{\omega}(f)|^{\gamma} = |Y_i{\omega}(f)|^{\gamma} - {\alpha_i}{\delta_i}|\hat{N}_i{\omega}(f)|^{\gamma}  
\end{equation}
The architecture of the proposed subtraction block is shown in Fig.~\ref{fig6b}.

A multiplier block is used to combine the estimated noise spectrum and $\alpha$ which is depended on the signal to noise ratio. The subtraction block is used to get the enhanced signal spectrum. The final enhanced spectrum is in the output of the multiplexer. Our proposed architecture requires four subtraction block for magnitude subtraction and four for the phase subtraction as shown in Fig.~\ref{fig3}.

\subsection{Signal Reconstruction Block}
\begin{figure}
    \centering
    \begin{subfigure}[b]{0.3\textwidth}
        \centering
        \includegraphics[height= 8 cm, width=1.5\textwidth]{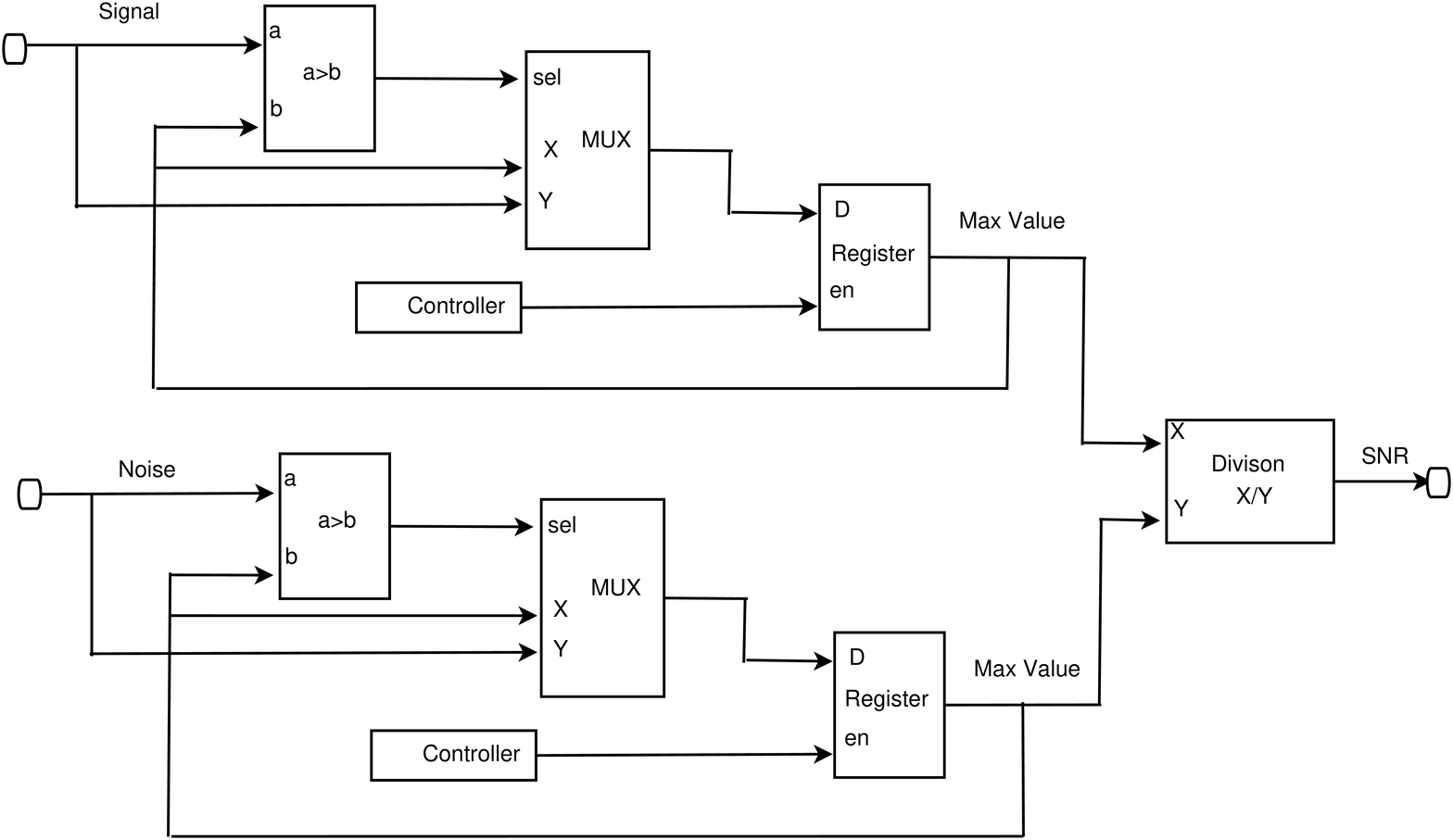}
        \caption{Signal to Noise Ratio Computation}
      \label{fig6a}
    \end{subfigure}
    \hfill
    \begin{subfigure}[b]{0.5\textwidth}
        \centering
        \includegraphics[height= 7 cm, width=1\textwidth]{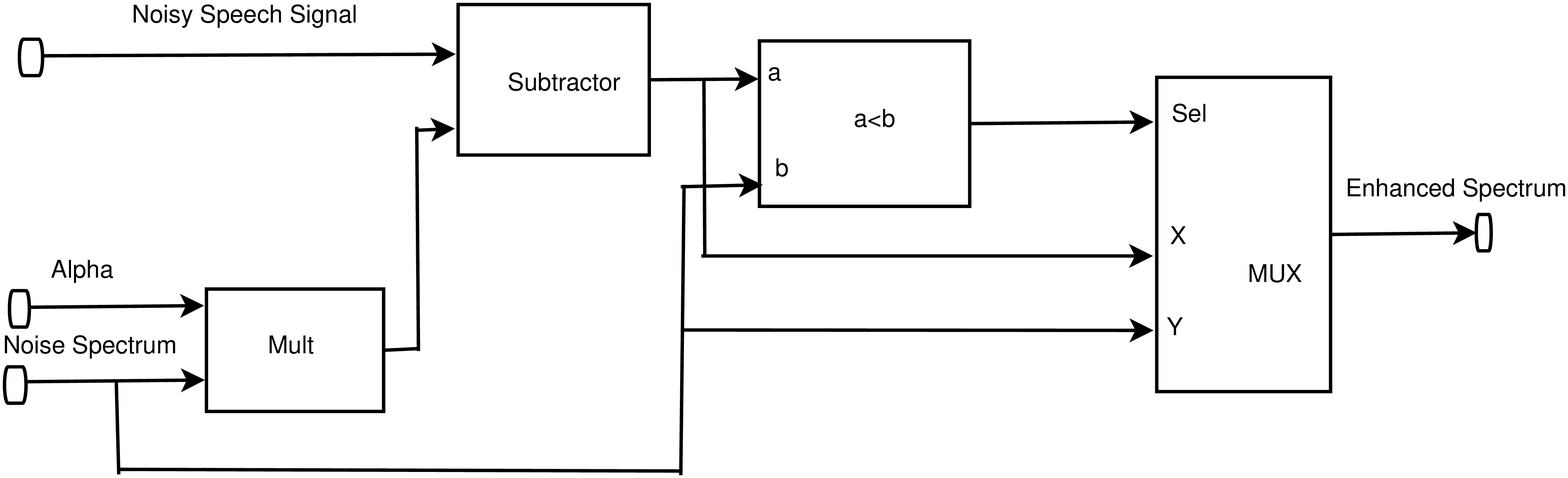}
        \caption{Subtraction Block}
       \label{fig6b}
    \end{subfigure}
 \hfill
    \begin{subfigure}[b]{0.4\textwidth}
        \centering
        \includegraphics[height= 7 cm, width=1.5\textwidth]{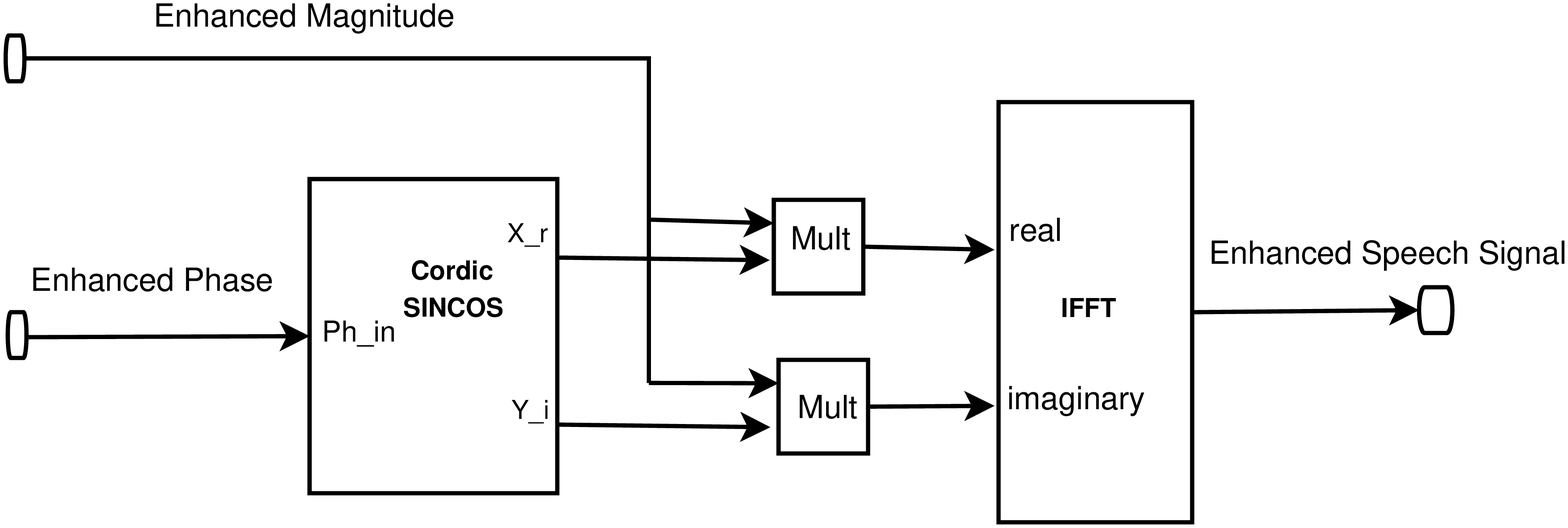}
        \caption{Inverse Fourier Transform Block}
       \label{fig6c}
    \end{subfigure}
 \caption{Block Diagram of Signal to Noise Ratio Computation, Subtraction and Inverse Fourier Transform Block}
    \label{fig5}
\end{figure}

Enhanced magnitude and phase spectrum are combined together to reconstruct the signal. Reconstruction of signal requires the inverse fourier transform over the magnitude and phase spectrum. The enhanced phase spectrum of the signal pass through the CORDIC SINCOS block~\cite{sysgen} to get their real and imaginary format. Two multiplier blocks are used to multiply the enhanced magnitude spectrum of the real and imaginary enhanced phase spectrum. The IFFT block used to reconstruct the signal and provide the enhanced speech signal. The reconstruction process of the signal is demonstrate in Fig.~\ref{fig6c}.

IFFT block is used to perform the inverse fourier transform operation over the signal. The option input/output was chosen for the IFFT to implement its pipelined versions. For the performance optimization of the FFT block, 4 multiplier structures are used and the phase factor is set to 8. Also the signal is segmented on non overlapping window of 256 samples. The architectural configuration of cordic SINCOS block is set to parallel mode for high throughput. The pipeline mode is set to maximum and the phase format is set to in the radians mode. The output width of this block is set to 16.

\section{Performance Analysis}
Field Programmable Gate Array (FPGA) contains a matrix of re-configurable logic circuitry. Different operations do not have to compete for the same processing resources because of the available special parallelism. So multiple control loops can run on a single FPGA device at different rates. The re-configurability of FPGAs can provide limitless flexibility. Most real-time systems require fast processing, which are met by the present day high speed FPGAs. The above mentioned hardware execution has been carried out on Atlys Spartan 6 FPGA board (Xilinx Spartan-6 LX45 FPGA, 324-pin BGA package,128Mbyte DDR2 16-bit wide data). Spartan-6 LX FPGAs are optimized for applications that require the absolute lowest cost. It provides up to 150K logic cells, integrated PCI express blocks, advanced memory support, 390MHz DSP slices, and 3.2 Gbps low-power transceivers.

\par 
Here, we have used all the sound sources from~\cite{sound_source1}~\cite{sound_source2} except market noise, railway platform noise and train horn noise shown in Table II. This market, railway platform and train horn noises has been recorded from the respective environment.  Football ground, market, car, railway platform, train horn and exhibition hall noisy signal was sampled at 16000 Hz and white, pink, cockpit, wind and factory noise are sampled at 8000 Hz. Due to the parallel nature of the proposed architecture and avoid sequential execution in software platform, we are implemented our design in FPGA. To compare our design with existing works~\cite{Boll:spectral_subtraction}~\cite{kamnath}~\cite{Zhang:real_imaginary},we have implemented ~\cite{Boll:spectral_subtraction}~\cite{kamnath}~\cite{Zhang:real_imaginary} in FPGA because no such hardware implementation was found in the respective literature.
\par
\begin{figure}
    \centering
    \begin{subfigure}[b]{0.3\textwidth}
        \centering
        \includegraphics[height= 7 cm, width=1.5\textwidth]{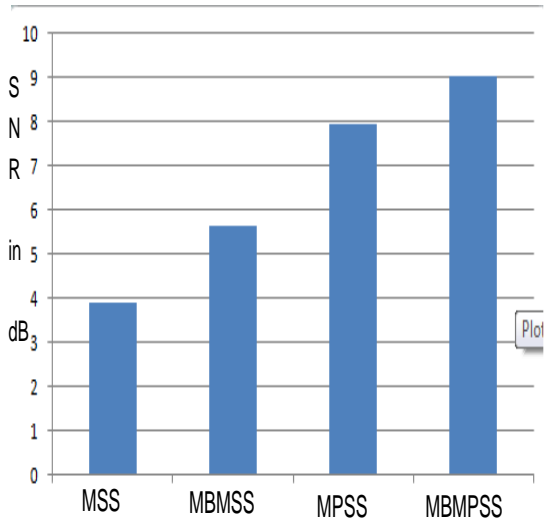}
        \caption{Rating of MSS, MBMSS, MPSS, MBMPSS}
      \label{fig11a}
    \end{subfigure}
    \hfill
    \begin{subfigure}[b]{0.4\textwidth}
        \centering
        \includegraphics[height= 8 cm, width=1\textwidth]{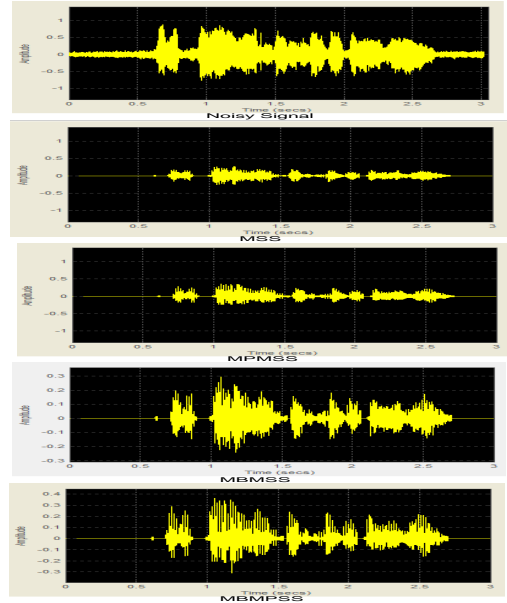}
        \caption{Scope representation of MSS, MPSS, MBMSS, MBMPSS}
       \label{fig11b}
    \end{subfigure}
 \caption{Rating And Scope Representation of MSS, MPSS, MBMSS, MBMPSS}
    \label{fig11}
\end{figure}

The device utilization of our implementation is shown in Table~\ref{tab:tab1}.. In Table~\ref{tab:tab3}, gives the system delay where only magnitude or phase operations are taking into account due to their parallel nature. Overall system unit delay of this design is $604$ and execute in  Xilinx Spartan-6 LX45 FPGA. The time requirement for execution of the proposed design on Xilinx Spartan-6 LX45 FPGA board is $6.04 microseconds$ where the board clock frequency is $100 MHz$. In Table~\ref{tab:tab2}, we compared the signal to noise ratio for all four methods from low SNR to high SNR. Rating of MSS, MBMSS, MPSS and MBMPSS are shown in Fig.~\ref{fig11a}. We observed that proposed method provides best result in every case in all SNR conditions. The melioration of the train horn noise was not outstripped due to high baseline. Time scope representation of the hardware implementation of MSS, MBMSS, MPSS and MBMPSS are shown in Fig.~\ref{fig11b}. So, we can conclude that our proposed design significantly outstrip the other existing methods in terms of SNR mainly.

\begin{table}[!hp]
\caption{Device utilization for SPARTAN 6 LX 45 FPGA} 
\centering
{
\begin{tabular}{|c| c| c| c|}
\hline\hline
Device utilization summary & Available &used &  utilization(\%) \\
\hline
Slice Registers &184,304&8451&4\\

Slice LUTs &92,152&7544&8\\

Slice  memory &21,680&1,161&5\\

Bonded IOBs & 296&42&14\\

DSP48A1S & 180&43&23\\
\hline
\label{tab:tab1}
\end{tabular}
}
\end{table}

\begin{table}[!hp]
\caption{SNR compression} 
\centering
\resizebox{13cm}{!}
{
\begin{tabular}{|c| c| c| c| c| c| }
\hline\hline
Input& input SNR  & MSS~\cite{Boll:spectral_subtraction} &  MBMSS~\cite{kamnath} & MPSS~\cite{Zhang:real_imaginary} & {\bf MBMPSS}(Proposed) \\
\hline
White Noise&-3 &0.53&1.79&3.95&5.01\\
         &0 &3.93&5.66&7.97&9.03\\
         &3 &6.99&8.01&11.07&12.25\\
         &8 &11.02&12.95&15.25&16.83\\
         &10&13.62&14.96&16.13&17.69\\
\hline
Pink Noise&-3 &3.10&4.58&7.63&8.91\\
     &0 &2.95&4.16&7.21&8.98\\
      &3 &5.59&7.13&10.01&11.28\\
     &8 &10.56&12.01&14.51&15.93\\
      &10&12.65&14.14&16.72&17.98\\
\hline
Cockpit Noise&-2 &0.67&2.35&5.08&6.16\\
            &0 &2.54&4.08&6.75&7.73\\
            &2 &4.32&5.23&7.78&8.26\\
            &6 &8.92&9.76&12.12&13.22\\
            &10 &12.91&13.96&16.83&17.99\\
\hline
Football Ground Noise &-5 &-1.11&0.02&2.88&4.01\\
                             &0 &3.64&4.84&6.89&7.81\\
                             &5 &8.95&10.09&12.27&13.12\\
                             &9 &12.25&13.39&16.25&17.34\\
                             &13&16.42&17.58&20.39&21.48\\
\hline
Exhibition Hall Noise&-3 &0.48&1.77&4.61&5.73\\ 
                         &0 &3.34&4.47&7.38&8.30\\
                         &3 &6.21&7.49&10.42&11.25\\
                         &8 &11.39&12.55&15.34&16.44\\
                        &12&15.42&16.56&19.51&20.47\\
\hline
Market Noise&-3 &0.42&1.60&4.44&5.56\\ 
                &0 &3.39&4.55&7.47&8.35\\
                 &3 &6.41&7.57&10.49&11.38\\
                  &8 &11.36&12.47&15.35&16.26\\
                &12&15.56&16.79&19.78&20.89\\
\hline
Railway Platform Noise&-5 &-1.08&0.15&3.01&3.92\\
                            &0 &3.43&4.69&7.56&8.49\\
                           &5 &8.35&9.47&12.38&13.26\\
                          &10 &13.46&14.55&17.51&18.38\\
                          &13 &16.39&17.52&20.30&21.11\\
\hline
Wind Noise&-3 &0.39&1.26&4.53&5.47\\ 
              &0 &3.47&4.64&7.32&8.61\\
              &3 &6.15&7.56&10.36&11.34\\
              &8 &11.46&12.43&15.27&16.36\\
              &13&16.37&17.43&20.28&21.07\\
\hline
Car Noise&-3 &0.49&1.76&4.53&5.61\\ 
            &0 &3.48&4.65&7.66&8.46\\
            &3 &6.54&7.69&10.52&11.49\\
            &8 &11.44&12.63&15.57&16.48\\
            &12&15.69&16.93&19.84&20.98\\
\hline
Factory Noise&-5 &-1.36&0.42&3.58&4.32\\
                  &0 &3.27&4.41&7.63&8.84\\
                  &5 &8.52&9.73&12.59&13.61\\
                  &10 &13.71&14.83&17.67&18.64\\
                  &13 &16.56&17.68&20.46&21.24\\
\hline
Bursting Noise&-3 &0.36&1.27&3.65&4.92\\
                  &0 &3.38&5.46&7.62&8.89\\
                  &3 &6.74&7.88&10.79&12.04\\
                  &8 &10.93&12.85&15.14&16.69\\
                  &10&13.47&14.71&15.93&17.42\\
\hline
Train Horn Noise&-3 &12.37&13.66&16.74&17.98\\ 
                     &0 &15.16&16.76&19.38&20.51\\
                     &3 &18.32&19.29&22.10&23.03\\
                     &8 &22.96&23.77&25.86&26.64\\
                     &12&25.74&26.13&28.75&29.43\\

\hline
\label{tab:tab2}
\end{tabular}
}
\end{table}  

\begin{table}[ht]
\caption{System Delay} 
\centering
{
\begin{tabular}{|c|c|}
\hline\hline
Hardware architecture & Delay \\
\hline
FFT &278\\
\hline
Magnitude-phase Extraction & 13\\
\hline
Magnitude/phase Operation & 24\\
\hline
Cordic SinCos & 11\\
\hline
IFFT &278\\
\hline
\label{tab:tab3}
\end{tabular}
}
\end{table}

\section{Conclusion}
In this paper, we have proposed a novel hardware design for speech enhancement based on the spectral subtraction algorithm. The subtraction procedure is performed on both magnitude and phase spectrum of the different frequency bands. In this way we are able to eliminate noise from high SNR signals as well as low SNR signals for the different frequency bands. FPGA based hardware implementation of the proposed architecture gives better performance in-terms of SNR and throughput over the existing architectures of MSS, MPSS, MBMSS.

\section*{Acknowledgment}
This work has been supported by the University Grant Commission (UGC) RGNF-2012-13-SC-WES-26014, Govt of India.

\end{document}